# On Calibrations using the Crab Nebula and Models of the Nebular X-ray Emission

M. C. Weisskopf[1], M. Guainazzi[2], K. Jahoda[3], N. Shaposhnikov[4], S. L. O'Dell[1],
V. E. Zavlin[5], C. Wilson-Hodge[1], R. F. Elsner[1]

[1] NASA Marshall Space Flight Center (MSFC), Space Science Office, Huntsville, AL 35812
[2] ESAC, P.O. Box 78, 28691 Villanueva de la Cañada, Madrid, Spain
[3] NASA Goddard Space Flight Center (GSFC), Greenbelt, MD 20771
[4] CRESST/GSFC, U. Maryland, Astronomy Department, College Park, MD 20742
[5] USRA/MSFC, Space Science Office, Huntsville, AL 35812

**Abstract**

Motivated by a paper (Kirsch et al. 2005) on possible use of the Crab Nebula as a standard candle for calibrating X-ray response functions, we examine consequences of intrinsic departures from a single (absorbed) power law upon such calibrations. We limit our analyses to three more modern X-ray instruments—the ROSAT/PSPC, the RXTE/PCA, and the XMM-*Newton*/EPIC-pn (burst mode). The results indicate a need to refine two of the three response functions studied. We are also able to distinguish between two current theoretical models for the system spectrum.

## 1   Introduction

The Crab Nebula, with its pulsar, exhibits remarkably complex morphology throughout the electromagnetic spectrum. The X-ray morphology is especially intriguing, showing a pulsar wind nebula with spectral shape dependent upon location. In view of spectral variations across the nebula and the necessity of joining the X-ray spectrum to a flatter spectrum at low energies and to a steeper spectrum at high energies, it is appropriate to ask, "How well does a single (absorbed) power-law fit the Crab Nebula's X-ray spectrum?" This question is particularly relevant in the context of a suggestion (Kirsch et al. 2005) that the Crab Nebula be used for calibrating X-ray response functions.

Figure 1 summarizes much of the data discussed by Kirsch et al. (2005) and other experiments which report absolute measurements of the normalization and power-law index of the Crab nebula plus pulsar emission. In the figure we limited our attention to those experiments for which an uncertainty in the power-law index is mentioned. Although the powerlaw index and normalization appear to vary, the fit to the straight line indicates to us that all these historical measurements are consistent with a constant flux, thus supporting the hypothesis that there may be a universal spectrum. *In many cases, however, the quality of the fit is not discussed*. Because of this, the uncertainties are quite possibly underestimated. We conclude, from the scatter of the data in Figure 1, that the power-law normalization, index, and thus flux are less precisely known than would be suggested by the statistical error bars from any single experiment. One (of many) possible systematic effects is the possibility that an absorbed power law is a poor description of the Crab's X-ray emission over a wide energy band.

We investigate this question from two directions. First, we analyze actual data sets for the Crab Nebula, obtained with three X-ray spectrometers, using the respective response matrix for each. Second, we simulate Crab data sets of the same size (2–6×$10^6$ counts) as the corresponding actual data sets, using two physically-motivated spectral models and the response matrix for each spectrometer. Then we fit those simulated data utilizing the same procedures as applied to the actual data.

For every data set—actual or simulated—we employ the $\chi^2$ statistic to test the null hypothesis that a single absorbed power law accounts for the data. In effect, the statistical analysis of the actual data probes both the adequacy of a power-law fit and the accuracy of the spectrometer's response matrix. Indeed, this is the basis for the suggestion that the Crab spectrum be used to calibrate the response matrix of X-ray spectrometers. The statistical analysis of the simulated data explores whether the model predictions would be distinguishable from a single (absorbed) power law, even if the response matrix were perfectly known. Observations with both XMM-Newton (Willingale et al. 2001) and with Chandra (Weisskopf et al. 2000) have shown that the value of the power-law index varies as a function of position in the Nebula. Since the sum of power laws is not a power law, our expectation is that the integrated Crab spectrum departs from a pure power law at some level. In this context, our analysis of the simulated data discussed below begins to quantify the ability of modern X-ray spectrometers to distinguish deviations from a power law given input spectra containing several million counts.

We analyzed data and performed simulations, for spectrometers aboard the *Röntgensatellit* ROSAT, the *Rossi X-ray Timing Explorer* RXTE, and the X-ray Multi-mirror Mission XMM-*Newton*. In particular, our study considered the following instruments and energy ranges:

1. ROSAT Position-Sensitive Proportional Counter (PSPC)—0.1–2.4 keV
2. XMM-*Newton* European Photon Imaging Camera (EPIC-pn) in "burst mode"—0.7–10.0 keV
3. RXTE Proportional Counter Array (PCA)—3.0–60 keV

We begin (§2) with a description of the actual data sets, analysis methods, and results. Next we discuss (§3) the simulated data sets, analysis methods, and results. Finally, we summarize (§4) the results and our conclusions.

## 2 Observations

Both in fitting and in simulating the X-ray spectral data, we utilized the XSPEC tool. For all spectral fits, we used the model of a single absorbed power law. For treating interstellar absorption, we follow Weisskopf et al. (2004), setting cross-sections to *vern* and abundances to *wilm* and using *tbvarabs*, which allows adjustment of the abundance of each element relative to the selected standard abundance—i.e., *wilm*. Such adjustment is necessary only for oxygen [O], which is under-abundant toward the Crab Nebula (Willingale et al. 2001; Weisskopf et al. 2004; however, c.f. Kaastra et al. 2009 which shows additional feature revealed at extremely high energy resolution) .

Here we describe the Crab Nebula observations and analyses for the ROSAT/PSPC (§2.1), for the XMM/EPIC-pn (§2.2), and for the RXTE/PCA (§2.3). Table 1 summarizes the results of fitting the observations with a single absorbed power law, for each of these X-ray spectrometers. Note that we quote errors on the best-fit parameters only if the fit is statistically acceptable: If the null hypothesis can be rejected, the formal statistical errors on the parameters are meaningless because the combined source model and response function do not describe the data. So-doing underestimates the errors.

For the multi-million count spectra considered here, the formal statistical errors are small. For example, the RXTE/PCA spectrum has over 6 million counts and the statistical errors on the spectral index are ±0.0013 (1 σ) or about 1% of the vertical span of Figure 1. If the null hypothesis can be rejected, however, then the formal statistical errors are of limited utility. In these cases, the statistical precision of the best-fit parameters could be estimated from simulation (using procedures similar to those described in section 3.1). However, the summary in Figure 1 already suggests that systematic errors dominate the uncertainties for two of the three satellites studied. The most likely cause of these "systematic errors" is a response matrix that is not sufficiently accurate.

## 2.1 ROSAT/PSPC observation

Following Kirsch et al. (2005), we analyzed ROSAT/PSPC observation 500065p, from 1991 March. Using the XSELECT package (version 2.4), we extracted the source+background spectrum from a 2.5′-radius circle centered on the pulsar and the background spectrum from a 4.5′–8.3′-radius annulus. We obtained the response function, pspcb_gain1_256.rsp (valid for data acquired before 1991 October 14), from the HEASARC archive[1]. Note that the Crab was not used to establish this response function: the PSPC was carefully calibrated on the ground (e.g. Briel et. al. 1988) and data from Markarian 421 was used to combine and adjust the measured detector response with the theoretical telescope response (G. Hasinger and F. Harberl, private communication. See also[2]). Adhering to the guide, *ROSAT data analysis using XSELECT and FTOOLS*[3], we prepared the source and background spectra from the event file's PI column, applying all standard corrections (for spatial and temporal gain variations, gain saturation, analog-to-digital non-linearity, and dead time).

Figure 2 shows the minimum-$\chi^2$ fit of an absorbed power law to the full (0.1–2.4 keV) ROSAT/PSPC data set. As Table 1 Case 1 documents, the quality of the fit is extremely poor. Inspection of Figure 2 indicates that most of the contribution to $\chi^2$ originates at the extremes of the energy range. Consequently, we re-fit the data over a restricted energy range (0.5–1.7 keV), displayed in Figure 3. Table 1 Case 2 confirms that the quality of fit is much improved; however, it is still statistically unacceptable. Nonetheless, the residuals are typically less than 1% of the total counts between 0.5 and 2.0 keV, where the vast majority of the counts are. To the extent that the emission from the Crab nebula is a smooth continuum, these residuals represent the maximum error in the response matrix. Although the response matrix is not "good enough" for million-count spectra, it is excellent over this reduced energy band.

Before we turn to the analysis of other satellite data, it is worth mentioning that our conclusion, that the Rosat Crab data are very poorly fit by a powerlaw spectrum, does not rely on the particular use of cross-sections, abundances, and treatment of absorption by interstellar dust grains. These might be considered to be relevant as the effects of the intervening medium are strongest in the energy band of the ROSAT/PSPC response---certainly when compared to the two other satellites we are examining. However, in Table 2 we list the number of degrees of freedom and the large, unacceptable values of $\chi^2$ for a number of other analyses of the ROSAT data. These analyses were performed using the XSPEC function *varabs* with cross-sections and abundances as indicated. The study agrees with the conclusion we drew from the ROSAT results in Table 1. The entries in Table 2 are from analysis when the amount of oxygen was left as a free parameter. Fixing the oxygen at its nominal relative abundance only serves to make the fits even worse, with $\chi^2$ increasing by a factor of 1.4 to 9.8.

---

[1] http://heasarc.gsfc.nasa.gov/docs/rosat/pspc_matrices.html
[2] http://heasarc.gsfc.nasa.gov/docs/rosat/ruh/handbook/
[3] http://heasarc.nasa.gov/docs/rosat/ros_xselect_guide/xselect_ftools.html

## 2.2 XMM/EPIC-pn (burst-mode) observation

Per Kirsch et al. (2005), we analyzed data taken with the EPIC-pn camera in "burst mode" with a position angle such that the full nebula lay on CCD #4. From observation #0160960401, we generated EPIC-pn event files using SAS 8.0 and the tool "epchain", with parameters identical to those in Kirsch et al. (2005). We employed calibration files available 2008 July and additionally corrected each event for the rate-dependent charge-transfer inefficiency[4]. We extracted the source+background spectrum from a 23-pixel wide region centered on RAWX column 35 (the column of maximum brightness), over RAWY rows 0–140, selecting only single and double events (with tags PATTERN = 0-4 and FLAG = 0). This data-selection region contains 92% of the events detected in rows 0–140. For background extraction, we chose the region RAWX = 1–8 and RAWY = 2–133. Choosing the CCD rows for extraction in this manner avoids spectral contamination from piled-up events in an artificial halo at higher RAWY (Kirsch et al. 2005). Figure 4 displays the burst-mode image and the selected extraction regions. Finally, to generate the response matrix and effective area files, we employed the tools "rmfgen" and "arfgen".

Calibration of EPIC-pn in Burst mode is partly based on, but does not fully rely upon, the Crab itself. The assumption that the Crab and several other sources have continuous smooth spectra in the 1.5-3 keV band was used to smooth out spectral variations at instrument edges in the correction for charge-transfer-efficiency. Moreover, the use of the Crab data have no bearing on broad band spectral parameters deduced from EPIC-pn in Burst mode observations. The calibration of this instrument mode is discussed in the references in the footnote.[5, 6]

Figure 5 shows the minimum-$\chi^2$ fit of an absorbed power law to the full (0.7–10 keV) XMM-*Newton*/EPIC-pn data set. (The burst-mode response is uncalibrated below 0.7 keV.) As Table 1 Case 3 documents, the quality of the fit is poor. Figure 6 exhibits the minimum-$\chi^2$ fit over a restricted energy range (1.0–7.0 keV), where we necessarily fix the relative abundance of oxygen because the instrument is insensitive to the oxygen in the line of sight. Table 1 Case 4 indicates little improvement in the quality of the fit. However, for either fit (Figure 5 or Figure 6), significant residuals occur near energies of known features in the instrumental response. Accordingly, we re-fit the full data set, excluding energies in the range 1.5–2.7 keV, which spans the Al-K, Si-K, and Au-M edges. As anticipated, the quality of the fit (Table 1 Case 3*) is much better although not excellent. Because we cannot reject the null hypothesis at 99.74% (3-$\sigma$) confidence, Table 1 gives the statistical errors in the best-fit parameters for this case (3*). Residuals in this fit are low but creep up to 5% in units of the ratio of the data to the model (see Figure 6).

Ideally, calibration of the EPIC-pn (or any other X-ray spectrometer) would rely upon a suite of standard candles with varied spectral properties. Unfortunately, due to the low duty-cycle (3%) required to avoid pile-up for the Crab, few X-ray sources can yield useful calibrations in this mode. Nonetheless, efforts continue toward extending the calibration base: For example 50 observations, about 30% of which targeted the Crab Nebula, contributed to the aforementioned rate-dependent charge-transfer-efficiency correction.

---

[4] http://xmm2.esac.esa.int/docs/documents/CAL-SRN-0248-1-1.ps.gz
[5] http://xmm2.esac.esa.int/docs/documents/CAL-TN-0018.pdf
[6] http://xmm2.esac.esa.int/docs/documents/CAL-TN-0083.pdf

## 2.3 RXTE/PCA observation

Again following Kirsch et al. (2005), we examined RXTE observation 50804-01-06-00, of 2000 December 16, limiting our analysis to data from Proportional Counter Unit 2 (PCU2). We employed HEAsoft version 6.4a[7], adhering to detailed step-by-step recipes[8]. In particular, we used filter files and good-time-intervals files from the standard products subdirectory, selecting data when the Crab was >10° above the horizon and <0.02° from the pointing direction, the PCU2 was operating, and the satellite was outside the South Atlantic Anomaly (SAA). Using PCABACKEST version 3.1, we estimated the background with the bright-source model pca_bkgd_cmbrightvle_eMv20051128.mdl. We extracted source+background and background spectra using SAEXTRCT version 4.2e, for all PCU2 xenon layers[9]. Next we corrected for dead time ($\approx$5%), using a standard recipe[10], and applied the correction to the exposure keyword in the source pha file. One of us (NS) generated the response matrix using the new energy-to-channel file pca_e2c_e05v04.fits and response-matrix generator released in FTOOLS 5.6.7 (2009 August 19) and the collimator response file p2coll_96jun05.fits from the CALDB database[11]. See Jahoda et al. (2006) and also Shaposhnikov et al. (2010) for more details about the response-function generation, calibration, and background model. In addition to including more data, Shaposhnikov et al. simultaneously fit for the energy-to-channel relationship, the quantum-efficiency parameters, and the redistribution parameters. This is an improvement, as the energy-to-channel parameters are coupled to the redistribution parameters through the partial-charge parameterization. The calibration of the energy-to-channel relationship (gain) and redistribution parameters (including spectral resolution) relies primarily on spectral lines from an on-board radioactive (Americium-241) source[12].

Figure 7 shows the minimum-$\chi^2$ fit of an absorbed power law to the full (3–60 keV) RXTE/PCU2 data set; holding fixed the column density and relative oxygen abundance. As Table 1 Case 5 documents, the quality of the fit is excellent. Note that the PCA response matrices use frequent observations of the Crab Nebula both to monitor time-dependent changes in the response and to tune the parameters of the physically motivated response model. Jahoda et al. (2006) assumed that the Crab Nebula plus spectrum can be approximated by a 2.1 photon-index power law. Now, however, Shaposhnikov et al. (2010) fit for the power-law index as well as the response parameters, yet find a similar result. They assumed an unabsorbed flux $2.4\times10^{-8}$ erg cm$^{-2}$ s$^{-1}$ (2-10 keV), at the higher end of many reported values. However, this does not significantly affect the present investigation, which primarily concerns the spectral shape. The assumption does impact the accuracy of the overall normalization deduced from this instrument, a point we return to in section 4. While the PCA calibration assumes that the Crab spectrum can be approximated as a power law, and the parameters of the PCA response matrix are tuned to minimize $\chi^2$ subject to this assumption, we can still rule out some models of the Crab spectrum (see below). Tunable parameters in the PCA model cannot create *arbitrary* features, particularly those with complex structure away from atomic edges important to the PCA response matrix. Of course the slope of the powerlaw continuum was not assumed, but derived.

---

[7] http://heasarc.gsfc.nasa.gov/docs/software/lheasoft/
[8] http://heasarc.nasa.gov/docs/xte/recipes/cook_book.html
[9] This procedure overestimates the errors on the background spectrum. PCABACKEST reports an estimate based upon a simple model fit to several hundred kilo-seconds of blank-sky data. The statistical uncertainty of the background estimate is dominated by the uncertainty in the particle-rate proxy (VLE rate). SAEXTRCT, however, reports the error as the square root of the number of predicted counts per 16 second interval, which is always a much greater number. This difference is important for the analysis of faint sources. For bright sources such as the Crab, the errors (and chi-square) are completely dominated by the source counting statistics, so the over-estimate of the background errors is inconsequential
[10] http://heasarc.nasa.gov/docs/xte/recipes/pca_deadtime.html
[11] http://heasarc.gsfc.nasa.gov/FTP/caldb
[12] The Americium-241 source provides 6 distinct calibration lines (or line blends) between about 14 keV and 60 keV. See Table 5 of Jahoda et al. (2006) for a listing of the line centers and identifications.

## 3 Simulations

As examples, we consider two spectral models for the Crab Nebula. (For another, see Atoyan and Aharonian 1996.) The first (Zhang, Chen and Fang 2008), which we dub "Model Z", is geometrically simple, resulting in a smooth spectrum (Figure 8). The second (Volpi et al. 2008), which we call "Model V", is geometrically more complex, resulting in a less smooth spectrum. Figure 9 compares the two models "Z" and "V" over the 0.1–60 keV band examined here. First we describe the methodology (§3.1) for the simulations and analyses; next we discuss the results for the two spectral models—Model Z (§3.2) and Model V (§3.3).

### 3.1 Methodology

In order to facilitate the simulations, Li Zhang and Delia Volpi kindly provided us the respective model spectra on a finer scale than used in the original papers. Moreover, to avoid systematic effects incurred when the model binning is coarser than the response binning, we interpolated (in log–log space) the provided spectra to generate spectral models covering the energy range 0.001–100 keV in 10000 logarithmically spaced steps. We then converted each model spectrum into a "table model" in XSPEC.

To the intrinsic spectral models, we factored interstellar absorption consistent with XSPEC settings used in fitting the observations—specifically, *vern* cross-sections, *wilm* abundances, and *tbvarabs* absorption model. For the simulations, we use a column density $N_H = 0.42 \times 10^{22}$ cm$^{-2}$ and a relative oxygen abundance [O] = 0.676 (Willingale et al. 2001; Weisskopf et al. 2004). Note that our conclusions are not sensitive to these settings.

We performed simulations, with counting statistics, using the XSPEC feature *fakeit* for each absorbed table model. In doing so, we employed response matrices identical to those used in fitting actual data sets. Analogous to each of the three observations described above (§2 and Table 1), we generated 100 realizations for each of the two spectral models. We selected the number of photons for each simulation to reproduce, on average, the number of counts detected in the corresponding observation, thus effecting realizations of statistical precision comparable to that of the respective observation. (NB: Because *fakeit* obtains random number seeds from the computer's clock read to 1-s resolution, we inserted the UNIX command "sleep 1" into our multiple-simulation script to ensure statistically independent realizations.)

We then use XSPEC to fit the resultant spectrum of each realization, to an absorbed power law, in a manner identical to that used to fit the observations (§2). Specifically, we fit over energy ranges and freeze fitting parameters to match the observational cases summarized in Table 1. Finally, we compute the average and standard deviation of $\chi^2$ and of each fitting parameter for the 100 realizations of each simulation case. The goodness of fit thus indicates whether a given observational case could distinguish the model from a single (absorbed) power law—if the response matrix were perfectly known.

### 3.2  Simulations using Model Z

Table 3 summarizes the results of fitting the simulated data for Model Z (Zhang, Chen and Fang 2008). In every case, a single (absorbed) power law gives an acceptable fit to the model. Not surprisingly in view of its spectrum (Figure 9), the Model-Z spectrum is indistinguishable from a power law in each X-ray observation considered. Consequently, *if Model Z represents reality*, the poor fits in the observed cases (Table 1) must result from residual errors in calibration of the respective spectrometer's response matrix over the energy range considered.

From the simulations, we find that distinguishing the Model-Z spectrum from a single power law would require at least 100 times more data for Case 3 and for Case 5. For the other cases, even more data would be needed.

### 3.3  Simulations using Model V

Table 4 summarizes the results of fitting the simulated data for Model V (Volpi et al. 2008). These results differ from those for Model Z, in that Model V exhibits spectral structure (Figure 9) departing from a simple power law over the X-ray band. The simulations show that the ROSAT/PSPC (Table 1 Cases 1–2) observation could *not* detect this spectral structure—even if the response matrix were perfectly known. This is not surprising: Inspection of Figure 9 shows that over the ROSAT/PSPC energy bands, Model V is quite close to a power law—albeit, somewhat flatter than that of Model Z.

The situation is more interesting for the simulations analogous to the XMM-Newton/EPIC-pn (burst mode) (Table 1 Cases 3–4) observation. Analysis of simulated observations over the restricted energy range (1.0–7.0 keV, Table 4 Case 4) could *not* detect this spectral structure—even if the response matrix were perfectly known. However, the departure of Model V from a single (absorbed) power law over the full range (0.7–10 keV, Table 4 Case 3) would almost be discernible at the 3-$\sigma$ level—if the response matrix were perfectly known. To quantify this, we repeated the simulation with 2, 4, and 10 times the number of counts, distinguishing Model V from a power law at 4.4, 8, and 19 $\sigma$, respectively.

Owing to its energy band, the situation with the RXTE/PCA (Table 1 Cases 5) observation is much clearer. *If Model V represents reality*, then the simulations (Table 4 Case 5) demonstrate convincingly that the RXTE/PCA data could not be fit by a single (absorbed) power law. However, the (absorbed) power-law fits to the actual RXTE/PCA Crab data are excellent—dramatically better than the Model-V simulations would allow. This indicates that the X-ray spectral structure predicted by Model V is *not* present in the spectrum of the Crab Nebula.

### 4  Summary and Conclusions

We began our study motivated by the paper of Kirsch et al. (2005), which seeks to determine the "correct" parameters of the Crab Nebula's power-law spectrum, for use in calibrating X-ray satellites. We re-analyzed Crab observations (§2) obtained with three X-ray satellite/instrument combinations—namely, ROSAT/PSPC (§2.1), XMM-*Newton*/EPIC-pn burst-mode (§2.2), and RXTE/PCA (§2.3)—to test the hypothesis that a single (absorbed) power law fits the respective spectral data. Owing to the large (multi-million) number of counts in each data set, the data are of exceptional statistical precision. In principle, each data set thus provides a sensitive test of this hypothesis. In practice, however, each data set is also sensitive to systematic errors in the response matrix of the respective instrument. Thus, inadequacies either in the source spectral model or in the response matrix can produce a statistically unacceptable fit ("bad") to the data. Conversely, a statistically acceptable ("good") fit indicates—except in special cases—that the spectral model and response matrix are each adequate to account for the data.

In attempting to fit an absorbed power law to the three Crab observations, we obtained very different results (Table 1) for the three instruments. For the ROSAT/PSPC observation, a power law is statistically a very poor fit—namely, >140 $\sigma$ and ≈14 $\sigma$ for the full (0.1–2.4 keV, Figure 2) and restricted (0.5–1.7 keV, Figure 3) energy ranges, respectively. While statistically a poor fit, the residuals over the restricted energy range are nonetheless small (≈1%, Figure 3) and their $\chi^2$ contributions concentrated in a few energy channels. Outside the restricted energy range, the residuals are significantly larger (up to ≈10%, Figure 2). Nevertheless, in either case, the parameters derived for the Crab from the analysis of the Rosat data are thus not universal and the measurements indicate the necessity for refinements.

For the XMM/EPIC-pn burst-mode observation, a power law is again statistically a poor fit—namely, 8.6 $\sigma$ and 7.5 $\sigma$ for the full (0.7–10 keV, Figure 5) and restricted (1.0–7.0 keV, Figure 6) energy ranges, respectively. However, the largest residuals are ≤10% and confined to energies near known atomic-absorption features in the instrument's response. Excluding energies near these features in the instrumental response, the power-law fit is statistically marginally acceptable—2.5 $\sigma$—with residuals less than a few percent. This confidence level is the effect of the current systematic ubcwertainties (5%) of the EPIC/pn response in burst mode. More calibration work is necessary before high-precision parameters for the Crab can be derived.

For the RXTE/PCU2 observation, a power law is statistically a good fit to the data over the energy range 3.0–60 keV (Figure 7). The residuals are small (≤1%) below 30 keV, but somewhat larger (few percent) at higher energies. As already mentioned (§2.3), the calibration of the RXTE utilized the Crab spectrum to fix certain parameters in the response matrix (Shaposhnikov et al. 2010). Consequently, the response matrix, especially the normalization, is not totally independent of the RXTE Crab data we analyzed here. Nonetheless, tuning parameters of the PCA model cannot create nor remove *arbitrary* spectral deviations from a power law, especially away from atomic edges in the PCA response. Thus, these data serve as the best measurements and only high-precision measurements of the powerlaw index and the column in the 3.0-60.0 keV band. One should be cautious, however, in extrapolating out of this band, especially towards lower energies.

We next investigated the potential sensitivity of such observations to the spectral model, apart from systematic errors in the response matrix. To do this, we performed simulations (§3) mimicking the observations, using the same respective response matrix and analysis techniques (§3.1) as we used to analyze the actual observations (§2). We generated the simulated data (using the XSPEC *fakeit* feature) using two recent source spectral models: Over the X-ray band (Figure 9), one spectral model for the Crab Nebula ("Z" for Zhang et al. 2008) departs very little from a single power law; the other ("V" for Volpi et al. 2008) shows potentially noticeable deviations from a single power law. The simulations demonstrate that, even if the respective response matrices were perfectly known, none of the 3 observations could distinguish a single power law from the first spectral model ("Z", Table 3). Indeed, even with an order of magnitude more

RXTE/PCA data, we could not discriminate between this spectral model and a power law. However, for two orders of magnitude more RXTE/PCA or XMM/EPIC-pn data, departures of Model Z from a single power law may be possible.

On the other hand, if the second spectral model ("V", Table 4) were valid, a power law would be an extremely poor (>200-$\sigma$) fit to the RXTE/PCA data, a marginally acceptable (2.9-$\sigma$) fit to the (full-range) XMM/EPIC-pn burst-mode data, and an acceptable fit to the ROSAT/PSPC data. That a single power law provides an excellent fit to the RXTE/PCA observation argues against models, such as that of Volpi et al. (2008), which show significant structure deviating from a simple power-law (or slightly curved) spectrum.

To summarize our analyses of the Crab observations (see Table 1), a single (absorbed) power law fits the RXTE/PCA data (Case 5) and (marginally) the XMM/EPIC-pn burst-mode data (away from known features in its instrumental response—Case 3*). The weighted average power-law index for these two observations is $\Gamma = 2.1020\pm0.0016$ which may be considered as a candidate for a universal parameter. There is only one (EPIC-pn burst-mode data—Case 3*), model-independent measure of the normalization, 9.210±0.053, that is not based on pre-conceived notions. These, together with the associated derived columns, may be considered as *candidates* for universal parameters. We emphasize that these are candidate parameters as known uncertainties in ground calibrations have not been explicitly accounted for. Thus, the uncertainties listed here should be considered as *lower* limits. For all other cases, we attribute the statistically significant departures from a power-law fit to inaccuracies in the respective response matrices: Typically these inaccuracies occur over limited spectral ranges—e.g., near known atomic edges—and are not large. However, they are quite noticeable owing to the high statistical precision of the multi-million-count Crab observations. In these cases, it appears appropriate to use the smooth spectrum of the Crab to tune response-matrix parameters to eliminate obvious artifacts—as was done for both the EPIC-pn burst mode (partially) and the RXTE/PCA response matrices. In addition to obvious artifacts of atomic edges, some response matrices—particularly, the ROSAT/PSPC—are clearly failing at the extremities of their spectral coverage. Here too, it seems fitting to tune response matrix parameters under the assumption that the Crab spectrum is a single (absorbed) power law. However, this approach can become problematic at low energies, owing to strong interstellar absorption.

## ACKNOWLEDGEMENTS

We thank Delia Volpi and Li Zhang for kindly providing us their respective models in a format that facilitated our analysis.

Table 1. Results of data analysis

| Case | Instrument | Band (keV) | Ct ($10^6$) | $\chi^2/\delta$ | $\Gamma$ | $N_H$ ($10^{22}$ cm$^{-2}$) | [O] | Norm |
|---|---|---|---|---|---|---|---|---|
| 1 | ROSAT/PSPC | 0.1–2.4 | 6.16 | 3343/227 | 2.04 | 0.43 | 0.43 | 9.37 |
| 2 | ROSAT/PSPC | 0.5–1.7 | 5.33 | 331/116 | 2.26 | 0.49 | 0.44 | 10.3 |
| 3 | XMM/EPIC-pn | 0.7–10 | 2.47 | 2386/1860 | 2.12 | 0.33 | 1.6 | 9.3 |
| 3* | XMM/EPIC-pn | 0.7–*–10 | 1.72 | 1760/1618 | 2.1075±0.0041 | 0.315±0.031 | 1.8±0.3 | 9.210±0.053 |
| 4 | XMM/EPIC-pn | 1.0–7.0 | 2.00 | 1568/1200 | 2.12 | 0.45 | ≡0.676 | 9.3 |
| 5 | RXTE/PCU2 | 3.0–60 | 6.66 | 94/86 | 2.1010±0.0017 | ≡0.42 | ≡0.676 | 10.938±0.016 |

Note 1: Uncertainties are given *only* if the fit is statistically acceptable—not rejected at the 3-σ level. When the fits are acceptable, the uncertainties are based upon one interesting parameter ($\chi^2_{min}+1$).

Note 2: Case 3* ignores energies in the range 1.5–2.7 keV, occupied by atomic features in the instrument's response.

Note 3: The normalization for Case 5 is not simply derived directly from the data but is based on an assumption discussed in the text.

Table 2. Results of analysis of Rosat/PSPC data with different models of the absorption and along the line of sight..

| abund | xsect | $\chi^2/\nu$ (0.1–2.4 keV) | $\chi^2/\nu$ (0.5–1.7 keV) |
|---|---|---|---|
| *angr* | *bcmc* | 4616/227 | 255/116 |
| *angr* | *obcm* | 4430/227 | 247/116 |
| *angr* | *vern* | 4608/227 | 232/116 |
| *wilm* | *bcmc* | 3421/227 | 308/116 |
| *wilm* | *obcm* | 3393/227 | 322/116 |
| *wilm* | *vern* | 3437/227 | 305/116 |

Note: The entries (*abund, xsect, angr, wilm, bcmc, obcm, vern*) represent their usual usage in XSPEC.

Table 3. Simulation results for Model Z

| Case | Instrument | Band (keV) | Ct ($10^6$) | $\chi^2/\delta$ | $\Gamma$ | $N_H$ ($10^{22}$ cm$^{-2}$) | [O] |
|---|---|---|---|---|---|---|---|
| 1 | ROSAT/PSPC | 0.1–2.4 | 6.16 | (228±21)/227 | 2.1921±0.0063 | 0.4210±0.0020 | 0.678±0.016 |
| 2 | ROSAT/PSPC | 0.5–1.7 | 5.32 | (116±17)/116 | 2.1914±0.0120 | 0.4209±0.0043 | 0.677±0.034 |
| 3 | XMM/EPIC-pn | 0.7–10 | 2.44 | (1877±62)/1860 | 2.1991±0.0026 | 0.440±0.019 | 0.61±0.10 |
| 4 | XMM/EPIC-pn | 1.0–7.0 | 1.94 | (1209±53)/1200 | 2.1966±0.0031 | 0.4242±0.0034 | ≡0.676 |
| 5 | RXTE/PCU2 | 3.0–60 | 6.66 | (89±13)/86 | 2.1958±0.0007 | ≡0.42 | ≡0.676 |

Note: Uncertainties are only given if the fit is statistically acceptable—not rejected at the 3-σ level. Uncertainties are based on the standard deviation of the scatter about the mean for 100 realizations.

Table 4. Simulation results for Model V

| Case | Instrument | Band (keV) | Ct ($10^6$) | $\chi^2/\delta$ | $\Gamma$ | $N_H$ ($10^{22}$ cm$^{-2}$) | [O] |
|---|---|---|---|---|---|---|---|
| 1 | ROSAT/PSPC | 0.1–2.4 | 6.16 | (229±22)/227 | 2.0701±0.0057 | 0.4214±0.0021 | 0.670±0.013 |
| 2 | ROSAT/PSPC | 0.5–1.7 | 5.32 | (117±17)/116 | 2.0680±0.0106 | 0.4234±0.0044 | 0.653±0.032 |
| 3 | XMM/EPIC-pn | 0.7–10 | 2.44 | (2038±74)/1860 | 2.1143±0.0023 | 0.524±0.020 | 0.29±0.08 |
| 4 | XMM/EPIC-pn | 1.0–7.0 | 1.97 | (1249±55)/1200 | 2.1078±0.0027 | 0.4476±0.0035 | ≡ 0.676 |
| 5 | RXTE/PCU2 | 3.0–60 | 6.66 | (3084±112)//86 | 2.22 | ≡ 0.42 | ≡ 0.676 |

Note 1: Uncertainties are only given if the fit is statistically acceptable —not rejected at the 3-σ level. Uncertainties are based on the standard deviation of the scatter about the mean for 100 realizations.

Note 2: RXTE/PCU2 data (Case 5) can easily distinguish Model V from a power law; XMM/EPIC-pn burst-mode data (full-range, Case 3) can nearly do so.

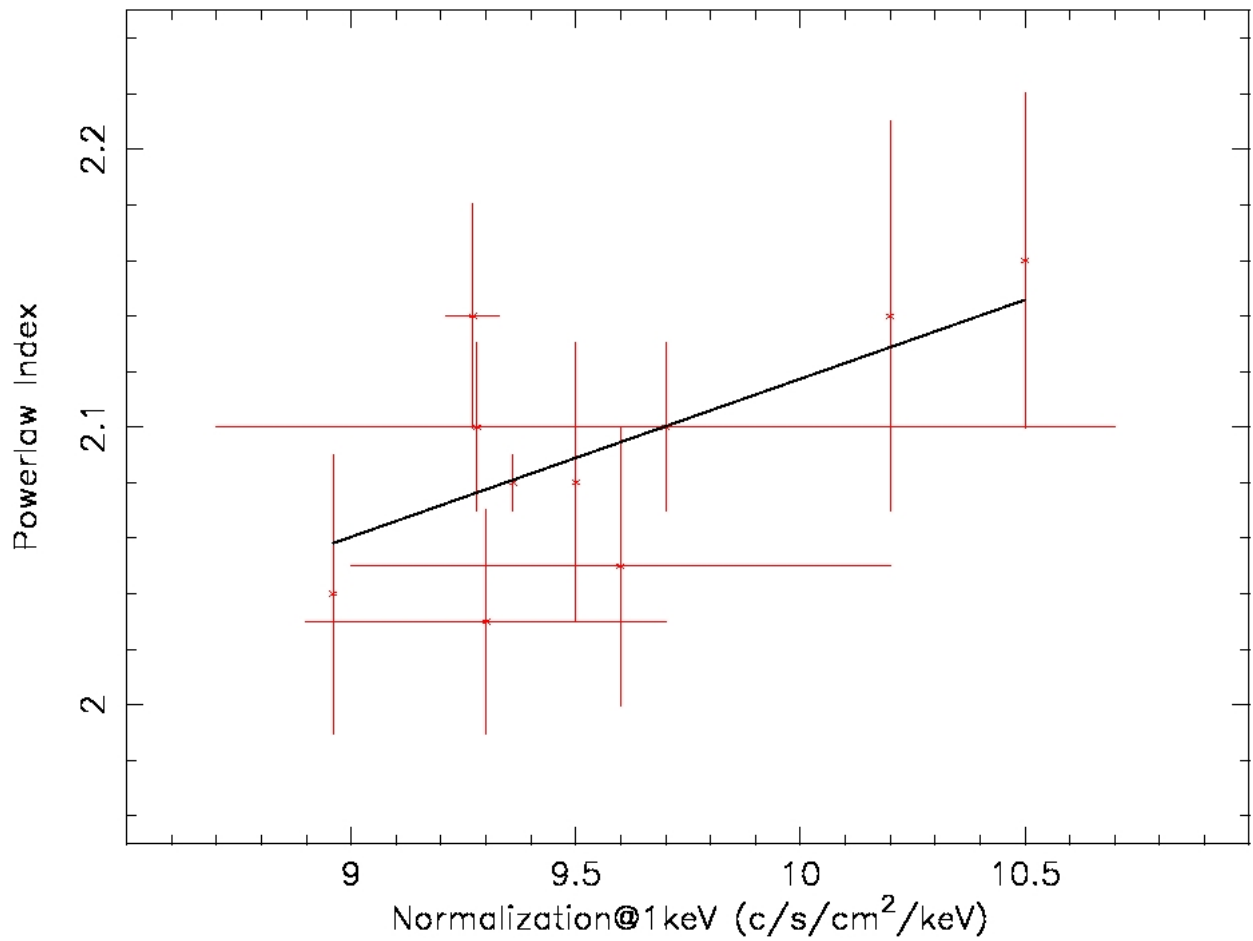

Figure 1. A partial summary of measurements of X-ray spectral properties of the Crab Nebula plus pulsar. Data points from left to right are as follows: Seward (1992), Kirsch et al. (2005), McCammon et al. (1983), two points from Koyama et al. (1984), Toor and Seward (1974), two points from Burrows (1982), and finally two additional data points from Seward (1992). We have not attempted to correct for various assumptions and values regarding the interstellar absorption. The line is a weighted best fit to the data.

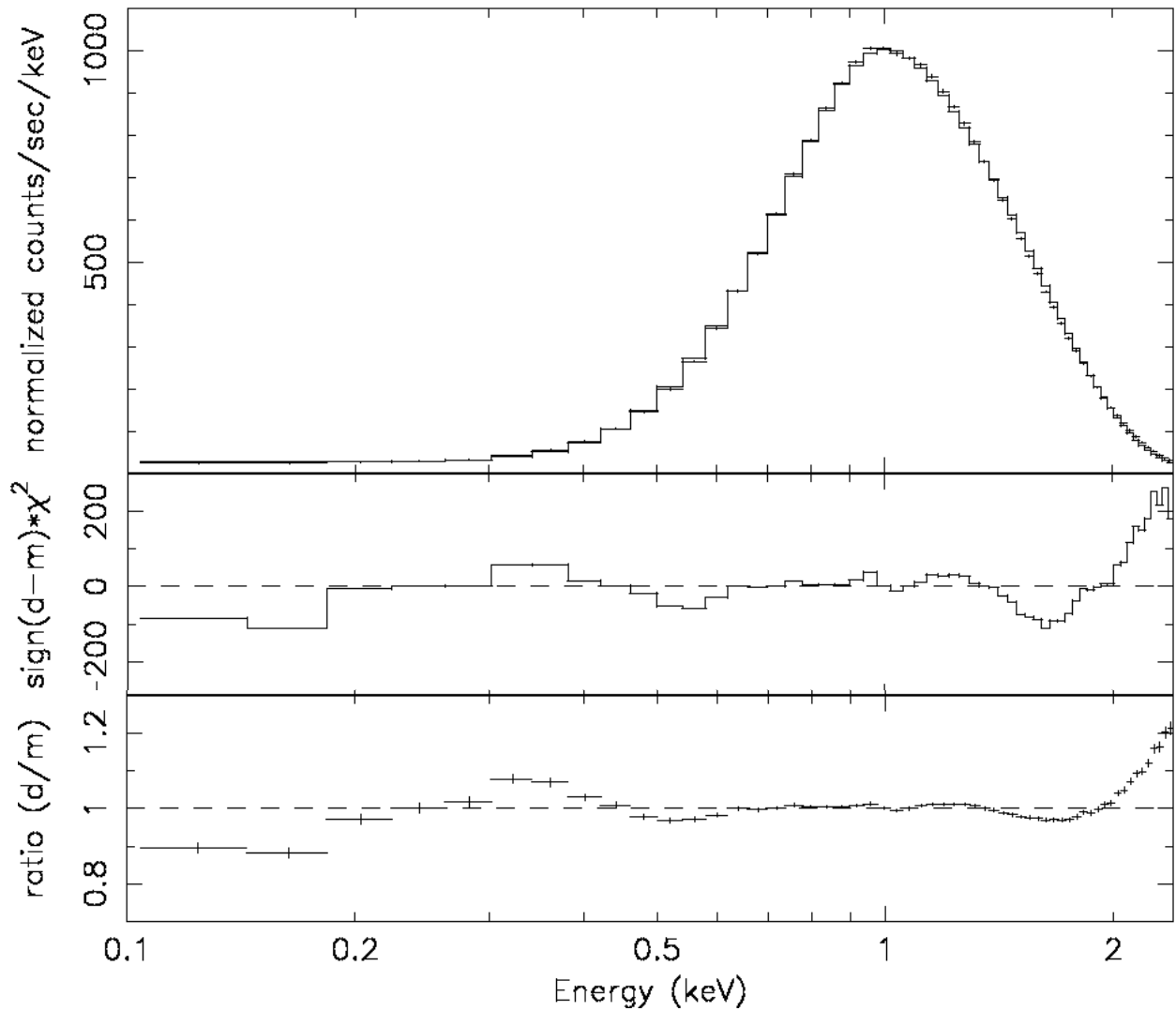

Figure 2. Best-fit absorbed power law for ROSAT/PSPC 0.1–2.4-keV data. The top panel compares data and model convolved with the instrumental response. The other panels show residuals, either as signed $\chi^2$ (middle) or as a ratio (bottom). For this fit, the column density, oxygen abundance, power-law index, and normalization are free parameters. The fit is quite poor ($\chi^2/\delta$ = 3343/227, Table 1), due largely to deviations at the extremes of the energy range.

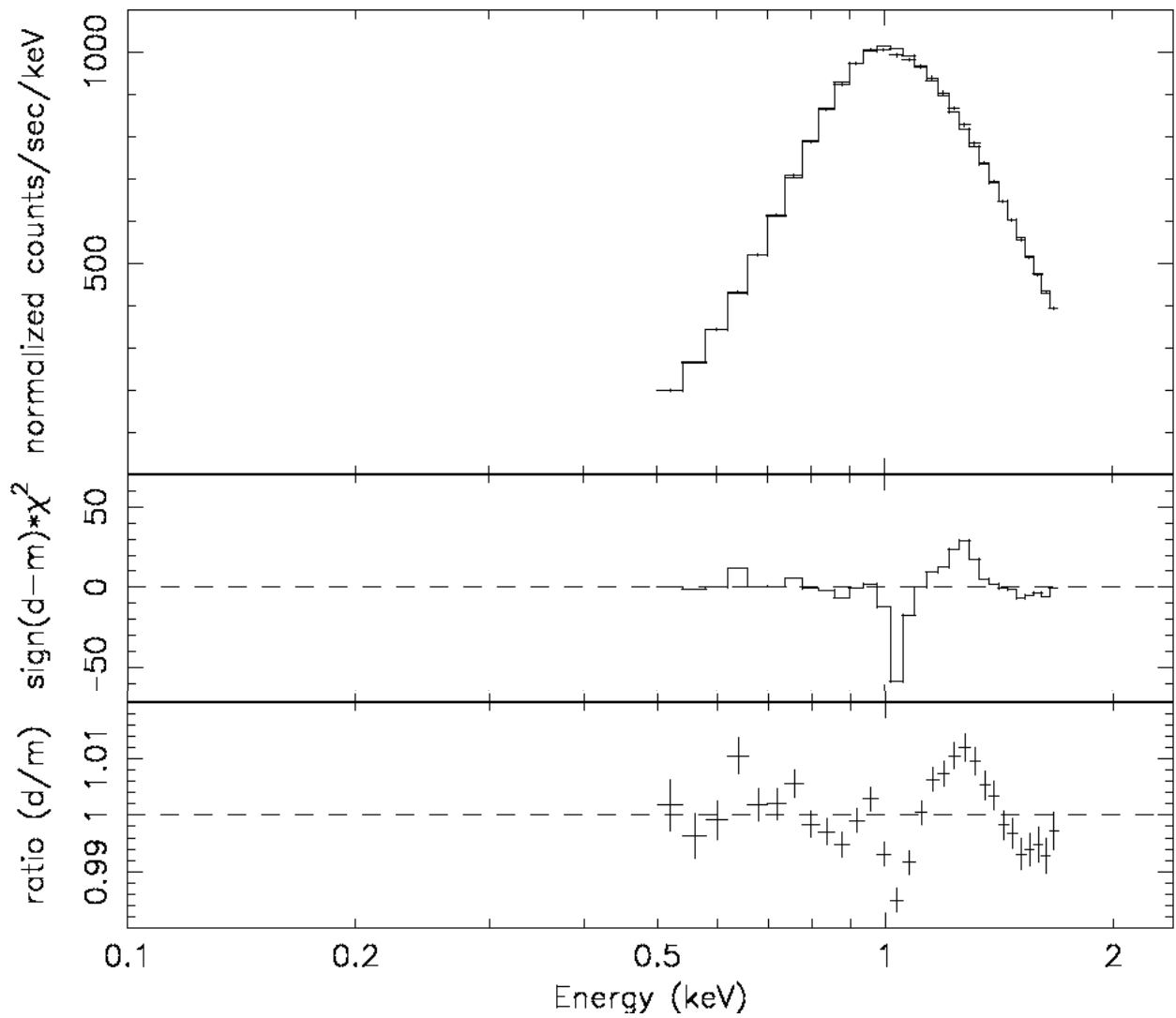

Figure 3. Same as Figure 2, but for ROSAT/PSPC 0.5–1.7-keV data. The fit over the restricted energy range is much better ($\chi^2/\delta$ = 331/116, Table 1) than that over the full range (Figure 2). Although the fit is formally (statistically) poor, the residual errors are small.

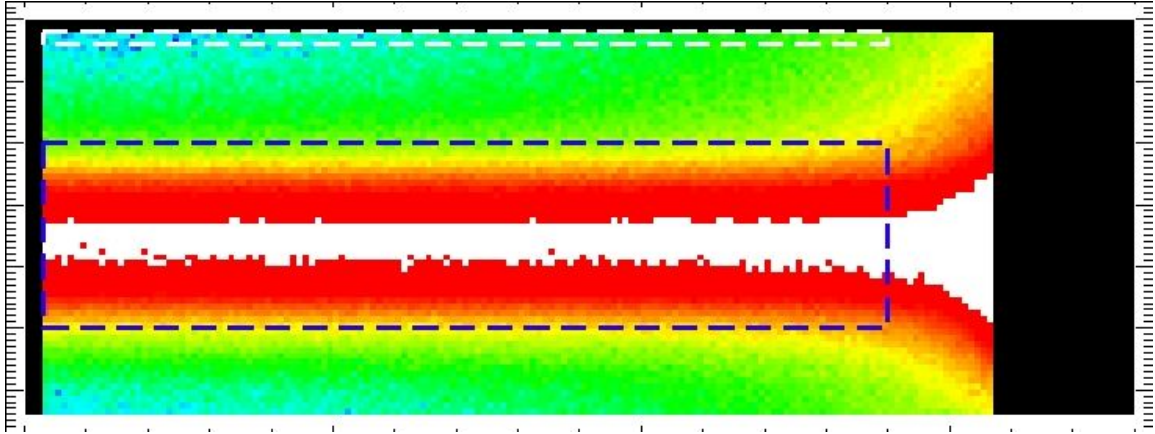

Figure 4. XMM-*Newton* EPIC-pn burst-mode image of the Crab Nebula. The blue dashed rectangle delineates the extraction region for the signal plus background; the white dashed rectangle, that for the background.

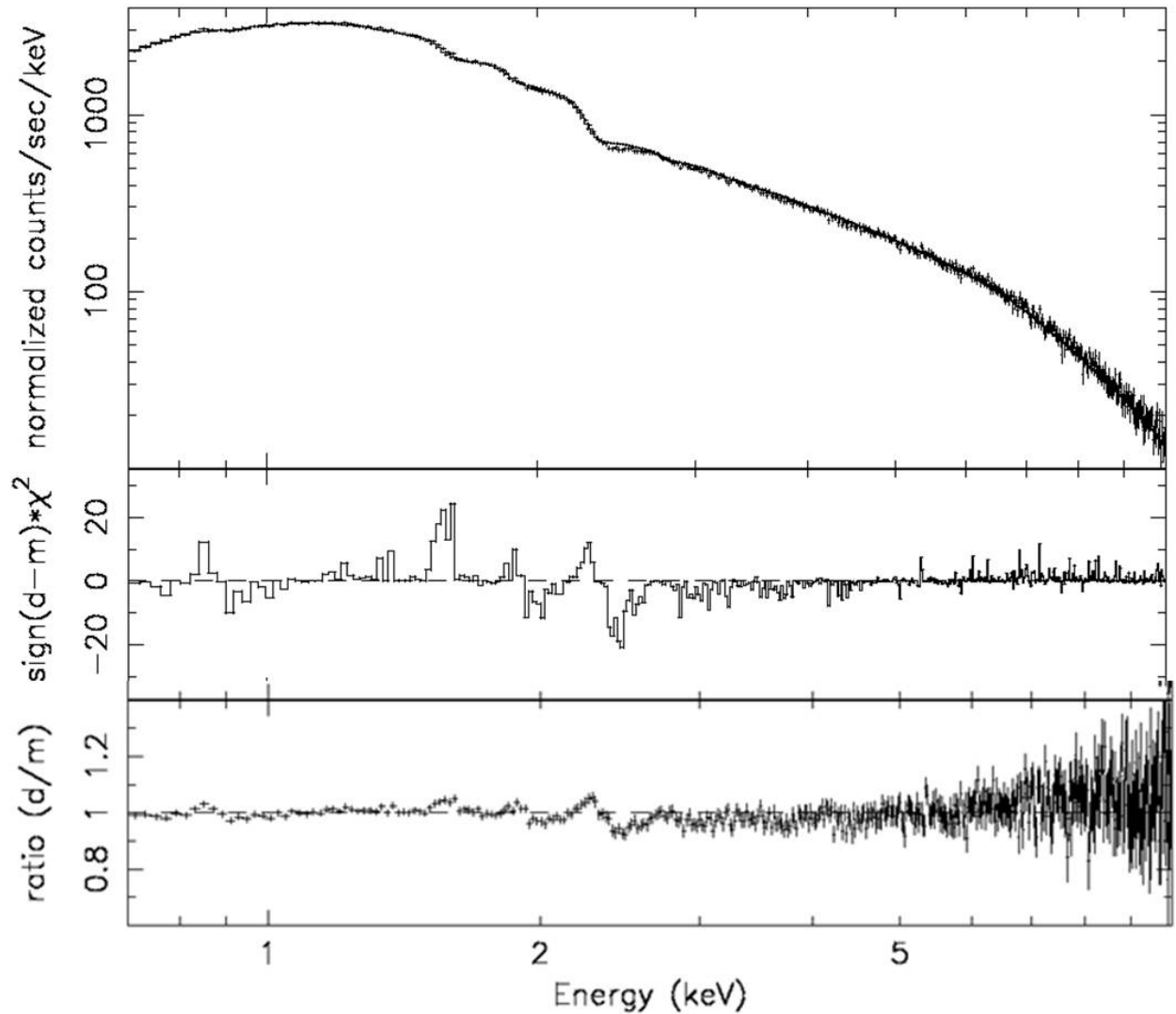

Figure 5. Best-fit absorbed power law for XMM-*Newton*/EPIC-pn burst-mode 0.7–10-keV data. The top panel compares data and model convolved with the instrumental response. The other panels show residuals, either as signed $\chi^2$ (middle) or as a ratio (bottom). For this fit, the column density, oxygen abundance, power-law index, and normalization are free parameters. The fit is formally (statistically) poor ($\chi^2/\delta$ =2386/1860, Table 1), due to deviations near known edges in the instrument's response.

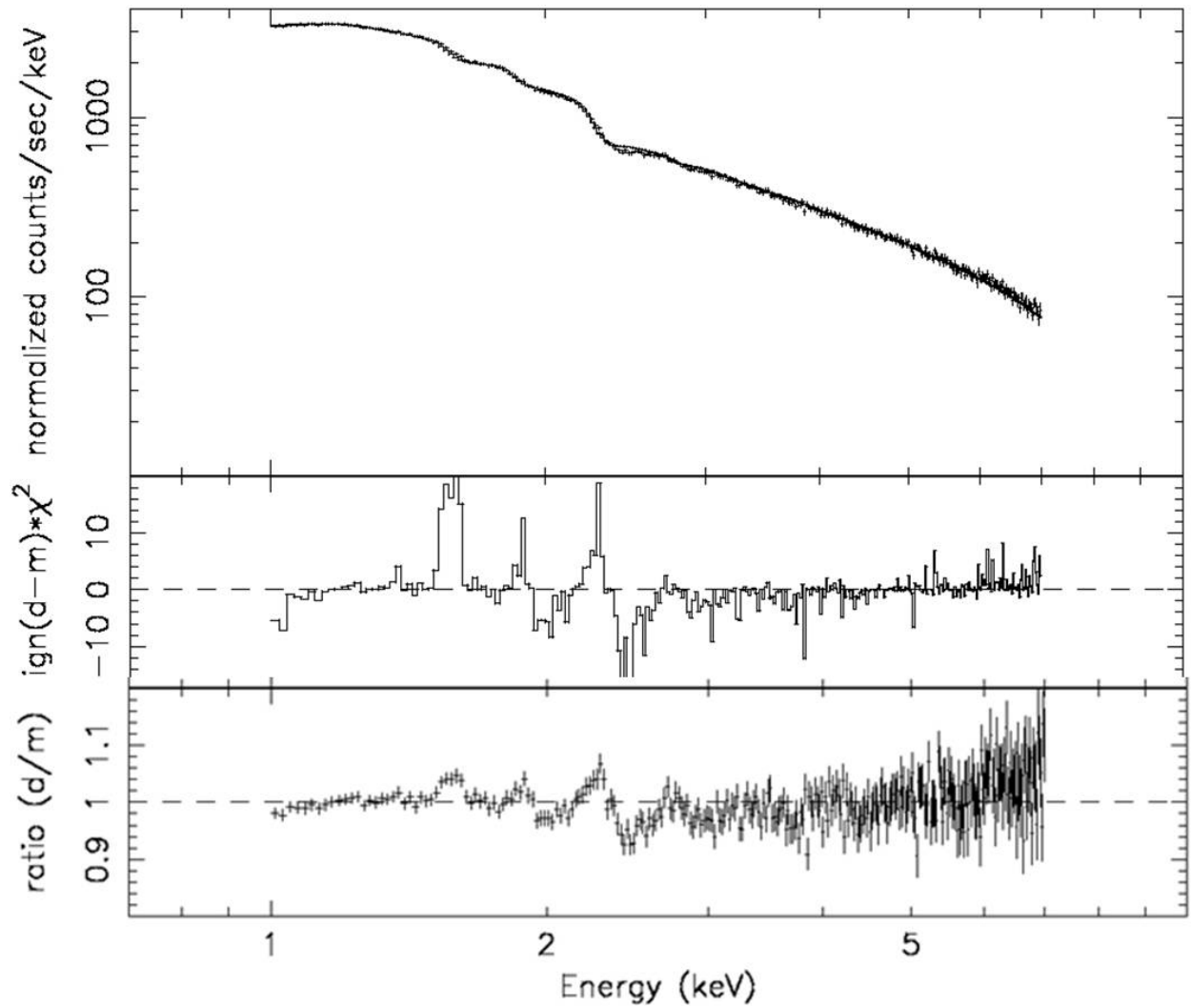

Figure 6. Same as Figure 5, but for XMM-*Newton*/EPIC-pn burst-mode 1.0–7.0-keV data. For this fit, the oxygen abundance is fixed (see text); the column density, power-law index, and normalization are free parameters. The fit over the restricted energy range is comparable ($\chi^2/\delta$ = 1568/1200 Table 1) to that over the full range (Figure 5). Thus, it is formally (statistically) poor, due primarily to deviations near known atomic edges in the instrument's response.

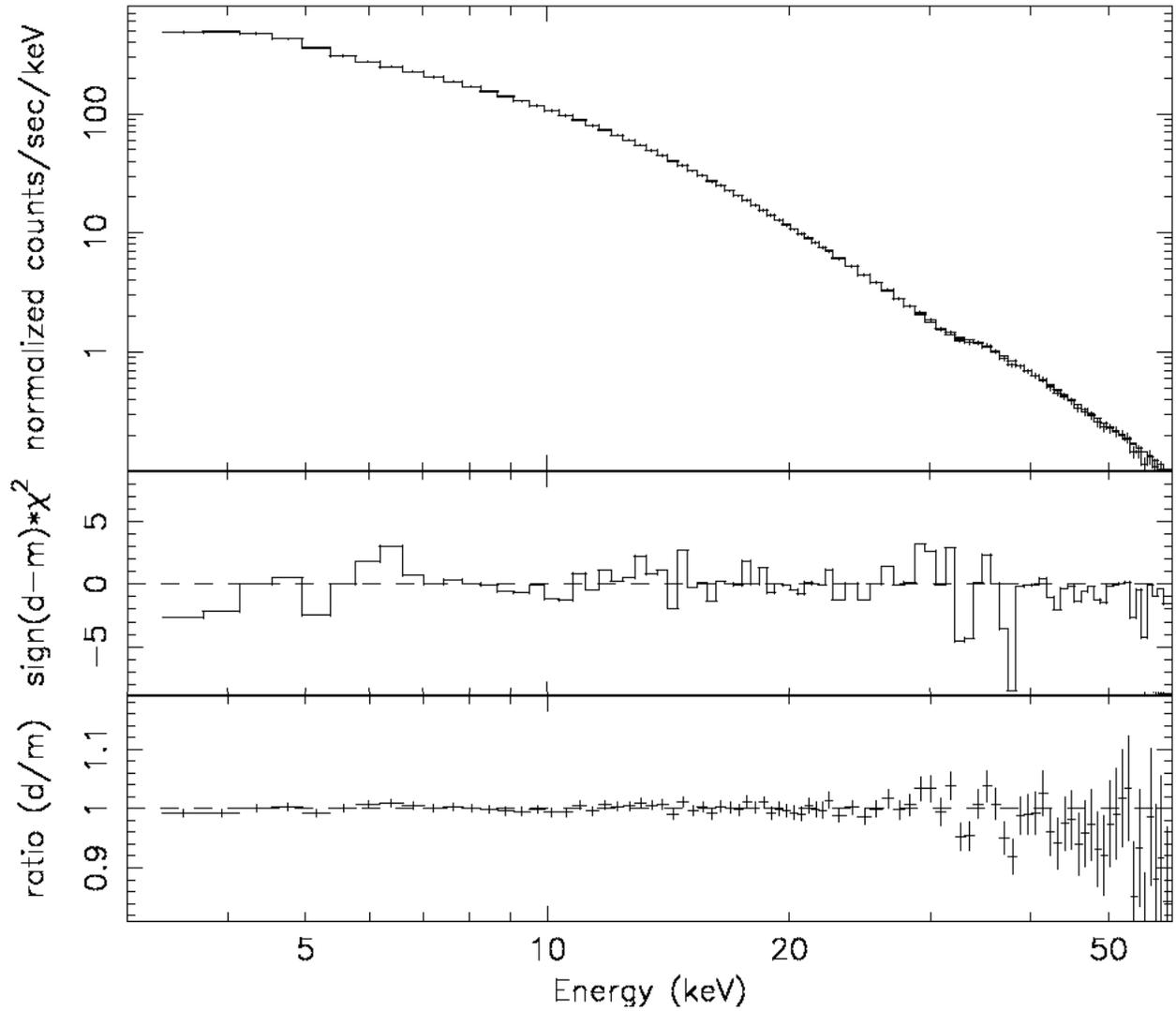

Figure 7. Best-fit absorbed power law for RXTE/PCU2 3.0–60-keV data. The top panel compares data and model convolved with the instrumental response. The other panels show residuals, either as signed $\chi^2$ (middle) or as a ratio (bottom). For this fit, the column density and oxygen abundance are fixed (see text); the power-law index and normalization are free parameters. The fit is excellent ($\chi^2/\delta$ =94/84, Table 1).

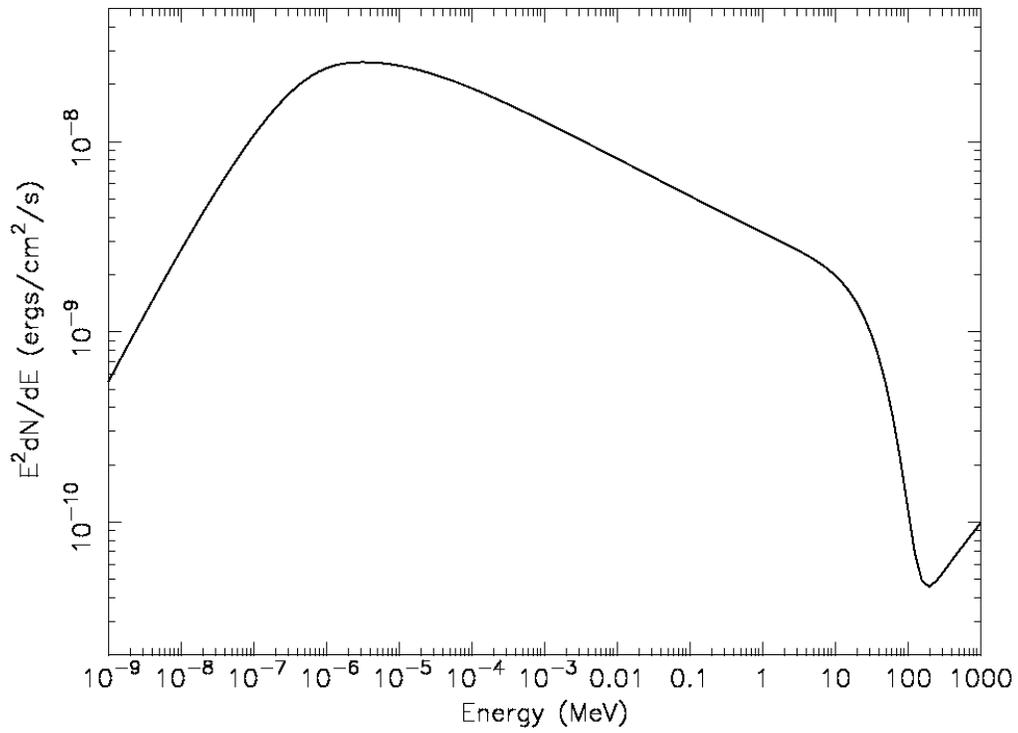

Figure 8. Intrinsic (unabsorbed) flux (E $F_E$ = ν $F_ν$) spectrum of the Crab Nebula, for the model of Zhang, Chen, and Fang 2008 ("Model Z"). Over the X-ray band, the spectrum differs little from a pure power law.

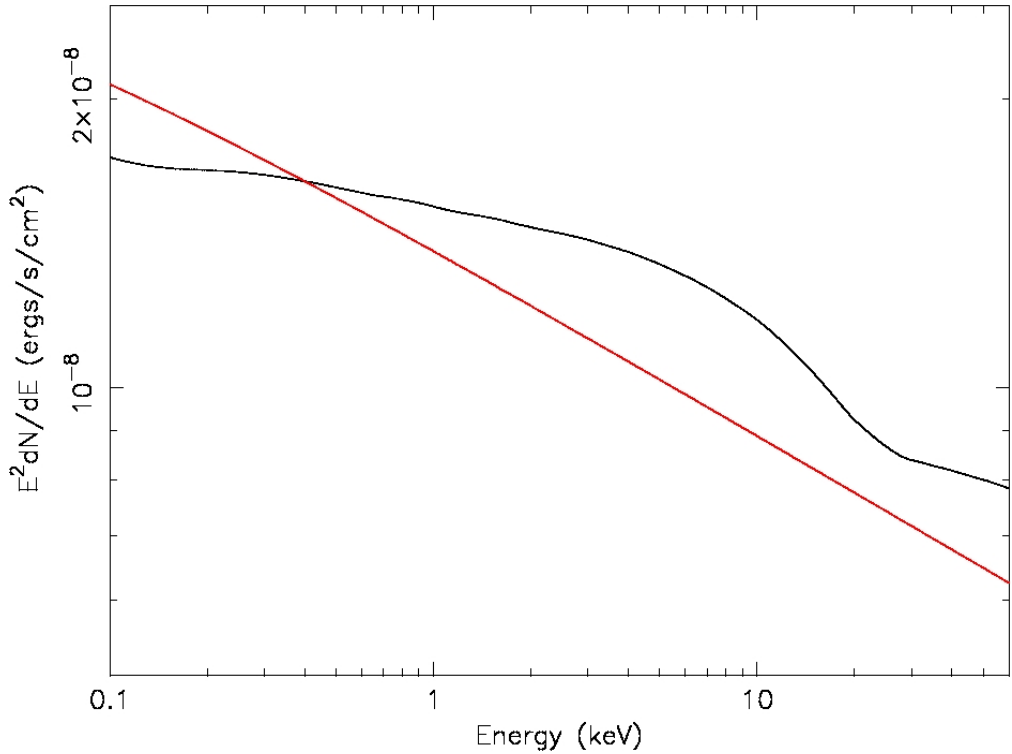

Figure 9. Intrinsic (unabsorbed) X-ray flux (E $F_E$ = $\nu F_\nu$) spectrum of the Crab Nebula, for the model of Volpi et al. 2008 (black line, "Model V") and for that of Zhang, Chen, and Feng 2008 (red line, "Model Z"). Model V has a couple kinks, both of which are in the RXTE/PCA band, and one of which is at the edge of the XMM-*Newton*/EPIC-pn band. The logarithmic bandwidths of EPIC and PCA are similar, so it is the location of the kinks that make the PCA more sensitive to the difference between Model V and a power-law model.